# UAV Command and Control, Navigation and Surveillance: A Review of Potential 5G and Satellite Systems

Nozhan Hosseini, Hosseinali Jamal,
David W. Matolak
Department of Electrical Engineering
University of South Carolina
Columbia, SC, USA
{nozhan@email., hjamal@email., matolak@}sc.edu

Jamal Haque
Honeywell Aerospace
Advanced Technology
CNS SATCOM & RF
Clearwater, FL, USA
Jamal.Haque@honeywell.com

Thomas Magesacher
CesiumAstro
Austin, TX
USA
thomas@cesiumastro.com

*Abstract*—Drones, unmanned aerial vehicles (UAVs), or unmanned aerial systems (UAS) are expected to be an important component of 5G/beyond 5G (B5G) communications. This includes their use within cellular architectures (5G UAVs), in which they can facilitate both wireless broadcast and point-to-point transmissions, usually using small UAS (sUAS). Allowing UAS to operate within airspace along with commercial, cargo, and other piloted aircraft will likely require dedicated and protected aviation spectrum—at least in the near term, while regulatory authorities adapt to their use. The command and control (C2), or control and non-payload communications (CNPC) link provides safety critical information for the control of the UAV both in terrestrial-based line of sight (LOS) conditions and in satellite communication links for so-called beyond LOS (BLOS) conditions. In this paper, we provide an overview of these CNPC links as they may be used in 5G and satellite systems by describing basic concepts and challenges. We review new entrant technologies that might be used for UAV C2 as well as for payload communication, such as millimeter wave (mmWave) systems, and also review navigation and surveillance challenges. A brief discussion of UAV-to-UAV communication and hardware issues are also provided.

### TABLE OF CONTENTS

## 1. INTRODUCTION

Unmanned aerial vehicles (UAVs), also known as unmanned aircraft systems (UAS) or drones, are being used for an expanding variety of applications. This ranges from consumer recreation flights, various military needs, crop monitoring, rail road inspection, etc., with aircraft sizes from several centimeters to several tens of meters. One critical need is to provide data connectivity for control and non-payload communication (CNPC), also known as command and control (C2) communication. The non-payload communication link is dedicated to secure and reliable communications between the remote pilot ground control station and the aircraft to ensure safe and effective UAV flight operation. This link can be either a line of sight (LOS) air-ground (AG) link between the two entities or a beyond-line-of-sight (BLOS) link using another platform such as a satellite or high altitude platform (HAP). Data rates for such links are expected to be modest (e.g., a maximum of 300 kbps for compressed video, which would not be used continuously).

In contrast, the payload communication link is usually used for data applications, and often requires high throughput. Payload communication types depend on application (e.g., agriculture, public safety), and can hence vary widely. The disruption of payload links—albeit inconvenient—is not critical, whereas CNPC link disruption can be critical. The functions of CNPC can be related to different types of information such as telecommand messages, non-payload telemetry data, support for navigation aids, air traffic control (ATC) voice relay, air traffic services data relay, target track data, airborne weather radar downlink data, non-payload video downlink data, etc. The cellular industry is also interested in using UAVs to expand their capacity and revenue to provide cost effective wireless connectivity for devices without coverage by the existing infrastructure. Additional cellular applications, e.g., as user equipment or relays, are also likely.

In this paper, we focus on the broader use of UAVs in the context of communications through ground infrastructure as well as satellite systems; see Figure 1. The disparate links (LOS and BLOS) mean different channel conditions and frequencies of operation, with different latency and range, and this increases challenges for the very high reliability required of CNPC links. In addition to existing cellular frequency bands (600 MHz to 6 GHz), the 5th generation (5G) cellular community is also considering the use of spectrum in the millimeter wave (mmWave) bands (24–86 GHz). In these bands, large free-space and tropospheric attenuations limit the link range, thus if the mmWave link is the only LOS link, when beyond the LOS mmWave range, BLOS capability will be needed. Such BLOS links are also of course required when in remote areas, out of range of any ground station (GS). Although satellites are an obvious choice for BLOS communications, the choice of satellite orbit, i.e., low-earth orbiting (LEO) or geosynchronous earth orbiting (GEO), distinctly affects the latency, link budget parameters, Doppler, and handoffs/handovers. In order to maximize frequency re-use, satellite operators are also planning use of narrower beams, which will increase handovers, further stressing connection reliability. Simply because of the much larger link distances, for currently planned BLOS frequency bands (above 5 GHz), to close the link between a UAV and satellite will very likely require the use of directional antennas and adaptively focused beams, i.e., mechanical or electronically steered antenna beams. Similar issues pertain to 5G mmWave links, but with far smaller antenna gain requirements. These issues of adaptive antennas, handovers and others complicate the system software and hardware, increasing the size, weight and power of the communication system.

Millimeter wave technologies that are likely to be deployed in 5G cellular systems bring large bandwidths for communication applications. The large available bandwidths allow fast commands, possibly transmission of local map data, etc. Also, mmWave links may be a case where C2 and payload communications are sent *together* on the same physical channel.

In addition to communications with UAVs, signaling must also be employed for navigation and surveillance. This is critical for safety, and can be challenging in some environments, such as very low-altitude flights near obstacles or remote areas. Thus, 5G use cases involving UAVs must also consider highly reliable navigation and surveillance methods.

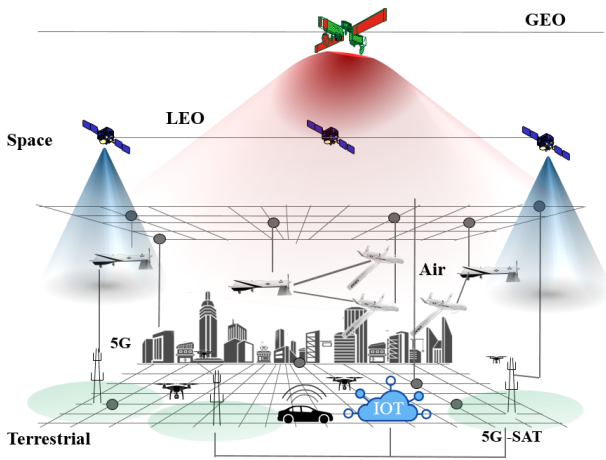

**Figure 1. UAV communication entities including ground, air, and space segments.**

The remainder of this paper is organized as follows: Section 2 briefly summarizes CNPC efforts. In Section 3, we discuss the satellite link for 5G UAVs, then in Section 4 we explain some of the potential new technologies that will likely be used in UAVs such as mmWave links. In Section 5, navigation and surveillance aspects of 5G UAVs are reviewed, Section 6 shortly discuss UAV-to-UAV communication and relaying in future potential cellular networks, and Section 7 is dedicated to prospective flight hardware architecture and trends. In Section 8 we conclude the paper.

## 2. OPTIONS FOR FUTURE CNPC

For medium and large UAVs, in the USA a standard has been adopted for CNPC, created by the Radio Technical Commission for Aeronautics (RTCA) [1]. This standard specifically pertains to the L-band (~900-1000 MHz) and a portion of C-band allocated to aviation (5.03-5.091 GHz). The standard applies to air-ground (AG) links (LOS) only, and the committee is at work on the BLOS standard. Estimated UAV CNPC bandwidth requirements for the year 2030 are 34 MHz for the terrestrial-based LOS CNPC, and 56 MHz for the satellite based BLOS CNPC link [2]. The RTCA standard does not specify any 5G applications, and primarily addresses the three lowest layers of the communications protocol stack. The standard is though general enough to be used for any type of 5G application involving medium and large sized UAVs.

In the United States, UAS CNPC deployment is planned in two phases, in which phase 1 supports terrestrial networks (based on proprietary handover functionality) but does not address any industry standard handover capability, which will be addressed in phase 2. The frequency bands allocated for CNPC phase 1 are L-band and C-band. For future CNPC BLOS CNPC, defined in phase 2, the satellite communications using L, C, Ku, or Ka bands, as well as networked terrestrial and C-band terrestrial will be considered.

For small UAVs, the situation is less developed. Both the RTCA and the cellular community (i.e., 3GPP) are conducting investigations for this use case, but work is still in progress. NASA also has a UAS traffic management program (UTM) that is working to develop air traffic rules and technologies for small UAVs at low altitudes, in coordination with the US Federal Aviation Administration (FAA). Multiple proof of concept field trials will be held in the coming two-three years. Physical and medium access control techniques will at least initially use commercial technologies such as cellular (LTE) and wireless local area networks (WiFi), but these are suboptimal in many UAV settings, over-designed for some CNPC links, and in addition are susceptible to jamming and spoofing. Hence work in this area is still continuing, and this can be a topic for 5G UAV research.

A potential candidate for terrestrial C2/CNPC-links for UAVs is the ultra-reliable and low latency communications

(URLLC) service category. This aims at an average latency of less than 0.5 ms and a probability of transmission success exceeding $1 - 10^{-5}$ through evolutionary and revolutionary changes in the air interface. One such interface is named 5G new radio (5G-NR) [3]. Recently developed dedicated short-range communications (DSRC) for vehicle-to-vehicle (V2V) communications in 5G [4] will also likely yield air-interface and network technologies that can be used or adapted to cover both AG links for individual UAVs as well as intra-swarm (air-to-air) links for UAV fleets.

Additional CNPC candidates include several proposed for so-called "long range" (LoRa) communications [5]. These are at least partly aimed at internet of things applications, and hence tend to support fairly low data rates (kbps) with simplified protocols. Reliability of links using these technologies would likely need improvement before they could be used for CNPC.

## 3. SATELLITE LINK CNPC

In many parts of the world, e.g., over the oceans, connections from UAVs to ground stations are difficult or even impossible. In those cases, BLOS, or, "beyond visual line of sight" (BVLOS) communications must be established. The use of satellite connections may be a complementary or required feature to improve or enable coverage and reliability both for commercial applications and for tactical missions.

Table 1 provides some comparative details on the three classes of satellite system orbits. Note that as the orbit altitude increases, so does latency: in the case of GEO, the latency can reach 0.5 seconds. This latency can strain the autonomous function of the UAV. Thus since the geostationary and medium earth orbit propagation delays may be impractical, the low earth orbit (LEO) constellations in 5G are gaining attention [6].

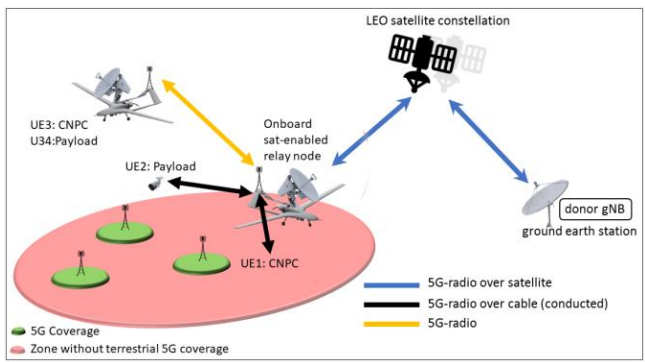

**Figure 2. BLOS CNPC link using the 5G relay node concept: onboard 5G relay node in the UA is connected to a donor NodeB in the ground earth station (GES) via a LEO constellation.**

The upcoming LEO satellite providers are focusing on providing data links, not just to ground users, but also to serve as the backbone for remote cellular towers, where cost of installing ground cables is prohibitive. Even though the propagation delay for LEOs is significant compared to that of terrestrial links, a single LEO satellite can provide a single-hop link between two points providing a footprint coverage of several hundred km.

**Table 1. Comparison of systems using different satellite orbits.**

| Satellite Type | LEO | MEO | GEO |
|---|---|---|---|
| Satellite height (km) | 500-1500 | 5000-12000 | 35800 |
| Orbital period | 95-115 min | 3-7 hours | 24 hours |
| # of satellites, coverage | 40-800, global | 8-20, global | 3, no polar coverage |
| Satellite life (years) | 3-7 | 10-15 | 15+ |
| handoff frequency | High | Low | None |
| Gateway cost | Very expensive | Expensive | Cheap |
| Doppler | High | Medium | Low |
| Round-trip propagation delay (UAV to control center and back via satellite link) | 10-30 ms | 70-200 ms | 0.5 s |
| Propagation path loss | Least | High | Highest |

A possible approach to smoothly integrate UAVs into the 5G systems via BLOS links using LEO satellites is the concept of 5G relay nodes (RN) [6] shown in Figure 2. A low-complexity satellite-enabled RN onboard a UAV transports the 5G downlink/uplink waveform via a LEO link to the actual base station, the so-called donor NodeB at the UA ground earth station. For the onboard equipment (flight controller requiring CNPC or payload equipment requiring a high-throughput link) the RN appears like a ground station. When flying in a swarm, the UAV with the satellite-enabled RN can act as a "cell tower" for the fleet. The donor NodeB, for whom the RN is transparent, simply sees a number of users. While this approach may not be optimal in the sense of achieving channel capacity, it requires less communication infrastructure both onboard the UAV and at the GS, simplifies handover from 5G to satellite, and leverages 5G technology.

## 4. POTENTIAL 5G TECHNOLOGIES FOR UAV CNPC

Cellular technologies are an obvious candidate for UAV CNPC links, but they have their shortcomings, as previously noted. The potential rapid maneuvering of UAVs and changes in antenna orientation may induce strong fading. Hence, modifications may be required in physical layer design, such as modifying orthogonal frequency division multiplexing (OFDM) (used in LTE) to filterbank multicarrier (FBMC), orthogonal chirp spread spectrum (OCSS), or some other modulation. The FBMC modulation is spectrally more compact than OFDM, so for cases in which the AG or satellite channels are non- or mildly-dispersive, FBMC could be attractive. In addition, traditional single-carrier modulations are also of interest for AG links, as they are of course also still used for most satellite communications today.

Another technology of note is multiple-input/multiple-output (MIMO) systems, realized by multiple antennas at both link ends. These are widely used in cellular and WLANs, but not yet in aviation (or in satellite links). Part of this is due to regulations for aircraft mechanical integrity, but size, weight, and power consumption constraints also pertain. As digital processing becomes more efficient and as frequencies get larger, MIMO for aviation applications will grow.

As noted, mmWave systems may be of use for UAVs. Because of the very large available bandwidths, in addition to CNPC communications, mmWave systems may also be used for *payload* communications. Highly directional beamforming antennas will enable much higher bit rates and aggregate capacity. A well-known concern regarding mmWave UAV links is the extremely high propagation path loss. Yet in practice these small wavelengths also enable greater antenna gain for the same physical antenna size. In mmWave UAV links, the time constraint for beamforming training will be more stringent than in static terrestrial mmWave communications due to UAV movement.

The high attenuation of mmWave signal blockage [7] is also a significant shortcoming. The use of mmWave links for UAVs may require precise flight algorithms to enable a UAV to avoid blockage zones and maintain LOS communication [8]. Yet in some locations such as built up areas where UAS can view entire streets or ascend above obstacles, UAS mmWave link reliability could be better than that of terrestrial links. Also as discussed and shown in [9]-[10] the reliability of mmWave links can be significantly improved with large antenna arrays that can increase directivity and reduce the co-channel interference for mmWave backhaul.

It is expected that terrestrial 5G systems will offer a minimum of 1 Gbps data rate "anywhere" to provide a uniform data rate experience to all users, and up to 5 and 50 Gbps for high mobility and pedestrian users, respectively [11]. The 28 and 38 GHz bands are currently available with spectrum allocations over 1 GHz. For these bands, based on channel measurement results in terrestrial environments [12], link range could be less than 200 meters for LOS cases and less than 100 meters for non-LOS (NLOS). These ranges highly depend on transmit power of course, which could easily increase for large UAV base stations in the future.

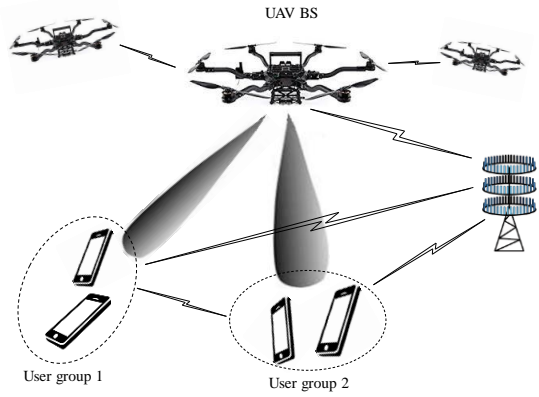

**Figure 3. UAV mmWave cellular communication concept and beamforming grouping technique for different user group.**

Comparing to terrestrial mmWave cellular networks, in a mmWave UAV network, it may be beneficial or even essential to use spatial-division multiple access (SDMA) or beam division multiple access (BDMA) [13]. The highly directional transmissions enable mobile stations in different locations to be separated into different groups using different spatial beams, as illustrated in Figure 3. Significant capacity improvement is possible, due to the large signal bandwidth and the use of SDMA [13]. The main challenge of SDMA is how the different users in different groups so called "group users" access the BS at the same time and frequency without interfering with each other. A practical strategy is to group users according to their angle of departures, where only users from different spatial groups can access the channel at the same time. Note that in UAV cellular, user grouping is not fixed due to UAV and ground users' mobility.

The use of mobile base stations as shown in Figure 3 can be an advantage for UAS since these BS can move to provide services to clients as needed. This has been proposed for users in rural areas and for emergency situations such as disaster relief for earthquakes and floods where the terrestrial ground BS are out of order. In high-speed vehicular ad-hoc (VANETs) and vehicle to everything (V2X) networks, cooperative data exchange and relaying using UAS may be an effective solution to the limited connection time of communication links between roadside units (RSUs) and vehicles.

## 5. NAVIGATION AND SURVEILLANCE FOR 5G UAVS

UAV integration into existing cellular infrastructure is being widely discussed in the literature, and 3GPP release 15, in draft now, will deliver the first set of 5G specifications. One of the focus areas of 5G is under the title "Critical Communications," and this includes "Drones & Robotics." Based on [8], 3GPP navigation and surveillance performance requirements for positioning accuracy in the 5G era will be 0.5 meter with 0.5 sec acquisition time in urban environments with a cell size on the order of 200 m. In the following subsections, we describe possible approaches for UAV navigation and surveillance methods for future 5G networks. Figure 4 depicts an overall view on proposed possible 5G UAV navigation and surveillance.

Numerous navigation and surveillance advances have recently occurred, in both literature and implementation. This includes improved inertial navigation systems, the use of more than one global navigation satellite system (GNSS), radar altimeters, LiDARs, and terrain databases. These are attaining precise performance in different military and civil applications.

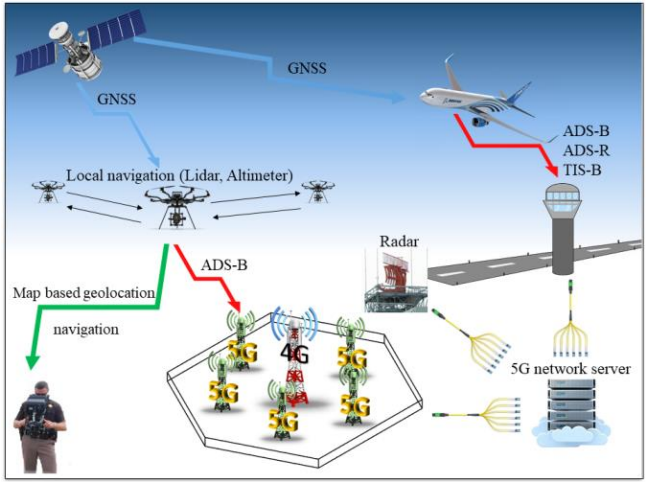

**Figure 4. Overall view on 5G UAV navigation and surveillance.**

*Global Navigation Satellite System (GNSS)*

Due to size, weight and power (SWaP) constraints, most UAS only use GNSS for primary navigation functions. However, some UAS are utilizing more accurate tightly-coupled embedded GPS/inertial navigation systems (INS) for navigation and platform stabilization. Standard GPS performs rather poorly in dense urban or indoor areas due to reduced satellite visibility. Moreover, lower accuracy can arise in case of jamming, spoofing and solar flares. Although disrupting GNSS signals is relatively easy and cheap, different techniques have been proposed to handle the interference impacts on the received signal, such as null filtering, cryptography, signal-distortion detection, direction-of-arrival sensing, etc. Research in [14] has found that a combination of different strategies can provide a reasonably secure countermeasure that could be commercially deployed.

*Automatic Dependent Surveillance-Broadcast*

Automatic Dependent Surveillance-Broadcast (ADS-B) is a surveillance technology in which an aircraft periodically broadcasts its position and receives messages from a ground station. ADS-B is the most popular technology to transform air traffic control away from radar-based systems [15]. The fast development of UAS has significantly outpaced their proper legislation, and current regulations prevent ADS-B out (transmitter) systems to be installed on unregistered systems (the vast majority of UAS). Due to not ensuring an acceptable level of air safety, surveillance regulations and proposals excluded small UAS in many documents [16]. Yet the range, resolution, accuracy and update rate in ADS-B are all superior to that of other existing technologies.

Integrating ADS-B into a standalone GNSS setup would enable operators (UAV pilots) to observe cooperative aircraft before they reach visual range. At present though, ADS-B bandwidth is only 1 MHz, and this severely limits its capacity. As the number of aircraft increases, this will make ADS-B inadequate, in areas where such information is most critical.

Another problem arises from the ADS-B limitation in supporting users in airspace below 400 ft (122 m). Almost all sUAS operate in very low level (VLL) airspace, defined by the FAA as under 500 ft (152 m) altitude. This issue might be addressed by 5G/B5G network technology in the near future, for example, by leveraging existing LTE systems for ADS-B [17]. However, limited coverage in rural areas and link initialization times are limiting factors in using only LTE. A cooperative approach between ADS-B and 5G UAV systems can alleviate the risk of co-channel interference on the 978 MHz ADS-B frequency channel.

In 5G networks, cellular footprints have evolved from macro BSs to small pico or nano BSs. More BSs in a given area with smaller footprints increases capacity and spatial spectral efficiency, yet this densification is achieved at the expense of increased handover rates. This may cause undesirable interruption in data flow in user equipment, and this can be a significant challenge in 5G UAV user navigation and surveillance applications. The authors of [18] proposed smart handover management schemes in single and two-tier downlink cellular networks as a solution to this problem.

*Radar*

According to [19], the minimum coverage value for a conventional air traffic control radar is 600 ft (183 m). The limited bandwidth or pulse repetition frequency provides aircraft position every 5 to 10 seconds. Newer equipment on the other hand may have minimum coverage capability as low as 100 feet (30.5 m). Radar accuracy degrades as aircraft size decreases, which makes the technology almost useless for small sized UAS under 500 feet (VLL). Another drawback of non-cooperative technologies such as radar and vision systems are their more complex hardware implementations and larger SWaP. State of the art waveforms and hardware design can overcome SWaP constraints. Reference [20], discussed an implemented frequency modulated continuous waveform (FMCW or chirp) sounder using low weight SDRs for UAV mounting.

*Map-based geolocation*

The term geolocation indicates a variety of techniques aimed at "mobility prediction," that is, tracking and computing the distance to or position of user or mobile terminals. Early research considered location determination techniques based on temporal measurements such as time of arrival (ToA), time difference of arrival, and enhanced observed time difference. These are all accurate in LOS scenarios whereas in urban environments, NLOS conditions with many MPCs will cause performance degradation.

In [21] a map-based location and tracking technique based on received signal strength indication was proposed for a GSM/3G network. This technique measures radio signal attenuation, assuming free space propagation of the signal and omnidirectional antennas. This method has a well-known triangulation position problem in which the accuracy of the estimated position strongly depends on the number of measurements and antenna placement. This paper also described an enhanced time forwarding tracking technique that exploits geographic information system map data and a predicted motion model to produce a set of candidate paths or shadow paths that improve map constraints. The large bandwidths available in 5G will enable an increase in delay resolution, and hence also in positioning accuracy for methods based on time of flight. In [22], an unscented Kalman filter based cascade solution was proposed for joint ToA and DoA estimation. Another expectation from 5G networks is to provide multiple antennas that enable direction of arrival and angle of arrival estimations [23].

*Local navigation (LiDAR, Altimeter and Dead Reckoning)*

In order to address the navigation challenges in indoor areas where GNSS signals are weak or blocked, light detection and ranging (LiDAR) has been proposed. LiDAR will work from infrared to the ultraviolet spectra, can measure distance, angle, velocity, vibration, posture and even shape of the object being illuminated. LiDAR utilizes a transmitter to emit a laser beam with known angle to the object, and by measuring the reflection of the laser return to the photosensitive transceiver sensor, distances can be calculated. Although LiDARs have been used by the military for many years, advances in design may make them suitable for commercial use. In [24], a LiDAR algorithm was proposed for navigation in robotic systems. Many startup companies are tackling the cost inefficiency of mass-producing LiDAR sensors. Commercial LiDAR systems in the market can range from $4,000 to $85,000 per unit. Black Forest Engineering, reduced the component cost from "tens of thousands of dollars" to $3 and expanded its production as reported in [25].

Dead reckoning is a navigation method in blind signal spots where the process of calculating UAS current position is based on a previously determined position, known and estimated speeds over elapsed time and heading at each speed. In [26], a non linear observer (NLO) and an exogenous Kalman filter (XKF) were used in an experiment where both estimators used the same set of IMU sensors (accelerometers, inclinometers and rate gyros), a camera and an altimeter. A machine vision system was employed to calculate UAV body-fixed linear velocity using optical flow. The results show that the inclusion of compensation for the additional biases reduced the position error of the NLO while the XKF can reduce the error even further by providing a better estimate of the velocity.

Another navigation tool for UAS is the radar altimeter, which in general is not sufficiently accurate. One proposed way to improve altimeter performance is to use Kalman filtering to obtain the optimum range estimate [27].

Addressing challenges in UAV navigation and surveillance systems may best be accomplished by merging several of the methods discussed here. Reference [28] integrated an inertial navigation system, GNSS and low-cost LiDAR to generate a high quality and dense point cloud data with 1 meter accuracy. Utilizing new radar navigation along with GNSS and ADS-B can be a promising solution for future UAS surveillance and navigation systems in 5G.

## 6.     UAV-to-UAV Communication

UAV to UAV communication (UUC) is a subcategory of air to air communication (AAC) and brings new challenges and opportunities. AAC links are crucial to evaluate CNPC link availability in any relay communications using multiple UAVs. Low level flying of UAVs means that proximity to earth surface objects will induce more multipath components (MPCs), and unless compensated for, will result in degradation of CNPC link availability. In [29], [20], the Rician amplitude fading model was found best for UUC communication links because it has a dominant LOS and multiple non-dominant NLOS paths. Reference [30] investigated wide-band channel models for airport parking and taxi environments, takeoff and landing situations, and en-route scenarios. UAVs on the other hand can have vertical takeoff and landing, and channel models for this have not been fully investigated. In [31], the authors used AAC models to evaluate the rate performance of a multiuser (MU)-MIMO configuration while proposing a mathematical framework for the analysis of several UAVs communicating in a mmWave frequency band with a central hub. A high relative velocity between UAVs will lead to large Doppler shift, and this requires more in-depth study, e.g., on appropriate waveform design for UAV AAC links. These waveform design studies should consider all AAC link challenges such as multi user interference (MUI), long ranges, channel distortions, and high velocities.

Some VLL altitude flying UAVs can benefit from occasional access to cellular networks on the ground, specifically for short-range 5G signals. In [32], a novel concept of three-dimensional (3D) cellular networks with polyhedron shaped cells was introduced, in effect integrating UAV base station and cellular connected UAV users. The authors' optimization algorithm showed significant reduction in latency and improved the spectral efficiency of the 3D UAV enabled cellular network compared to the classical SINR based network. For future 5G mmWave signals, their short-range limitation might be addressed with flying relay node BSs.

The prospective view on UUC considering all mentioned challenges is illustrated in Fig. 5.

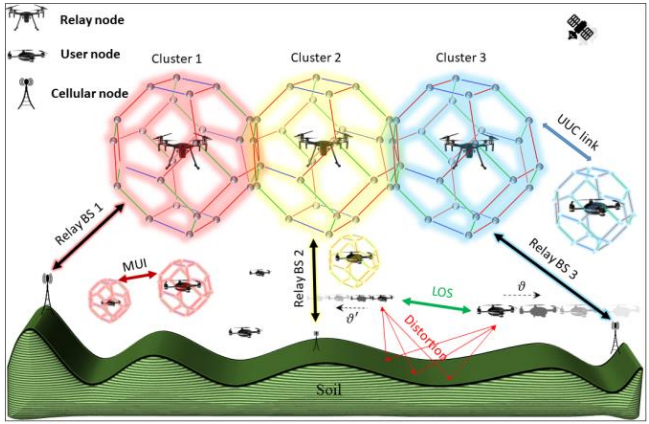

**Figure 5. Prospective view on UAV-to-UAV communication.**

## 7. Flight Hardware Architecture and Trends

As the market for airborne systems grows, design and manufacturing for both commercial and military UAVs tend towards lower cost and shorter design cycles. This often means adopting the commercial off the shelf (COTS) philosophy for flight hardware. Hand in hand with COTS goes the trend towards FPGA-based software-defined radio (SDR) platforms combined with single-board computers that leverage modular and model-based system design platforms (such as GNU radio) [33].

Hardware design for communications on UAVs is a tradeoff between several interrelated aspects and is typically governed by the following optimization criteria:

- SWaP-C elements
- Instantaneous bandwidth (IBW)
- Range (which is tightly coupled to available RF power, receiver sensitivity, antenna gains)
- Link reliability, integrity, and continuity.

Note that although not all criteria apply to CNPC directly (such as IBW), they are important to system design as a whole since the same hardware may also host payload processing.

Figure 5 shows an example of a flexible and programmable communication-system platform based on small-form-factor hardware modules [34]. The choice of system architecture can have a strong impact on CNPC. Small form-factor modules allow the system designer to improve the CNPC-link reliability by choosing the required amount of hardware redundancy. Programmability and re-configurability allow for a tight integration of the CNPC communications system and the actual system to be controlled (such as flight controller, navigation system, or surveillance system) on the same hardware platform (providing the appropriate amount of hardware redundancy). For example, application-specific onboard processing (such as angle-of-arrival detection for navigation or image feature extraction and compression) can be hosted on the same FPGA fabric the CNPC modem is running, which reduces latency and eliminates a chain of potential error sources introduced by interfaces and cable connections between separate hardware modules that would be necessary otherwise.

Advanced antenna systems can offer several advantages. For example, exploiting diversity over multiple antennas mitigates the impact of fading and thus improves link reliability. A more advanced example is forming steerable beams using antenna arrays, which results in a higher antenna again along the steering direction compared to an omnidirectional antenna. Furthermore, steerable beams at transmitter and receiver reduce the probability of the radio activity being detected and jammed, respectively, which may be of importance in certain applications. However, advanced antennas also introduce system-level challenges such as antenna-array calibration or the need for location and attitude information required for beam steering.

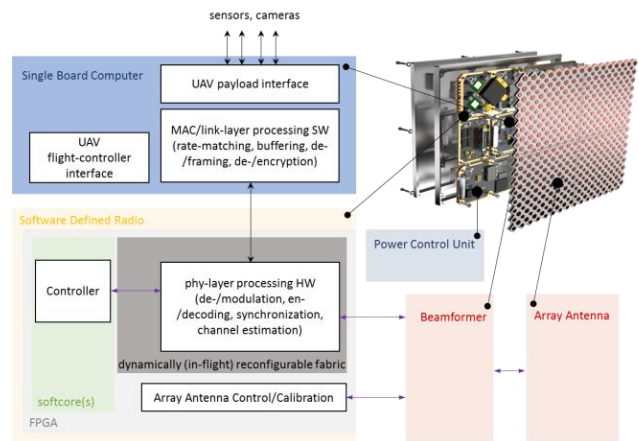

**Figure 5. SDR based flight-hardware architecture with beamforming Ku-band phased-array antenna in development at CesiumAstro, Inc. [34].**

## 8. CONCLUSION

In this paper, we reviewed UAV Command/Control (C2) (or, Control and Non Payload Communication, CNPC) links for both space and terrestrial systems. The integration of UAVs into 5G using both terrestrial and satellite links was discussed. We provided a short summary of existing standardization and research efforts, followed by a description of how satellite links might be used for these critical links. Potential future CNPC technologies for 5G UAVs were then described, with some emphasis on broadband mmWave systems for short range C2. Critical navigation and surveillance techniques were then reviewed. We noted that for the very high reliability required for C2 links, multiple integrated navigation technologies would likely be required. Our final section covered hardware challenges that will likely be addressed by state of the art flight hardware architectures with multiple high-performance SDRs and phased array antennas.

## BIOGRAPHY

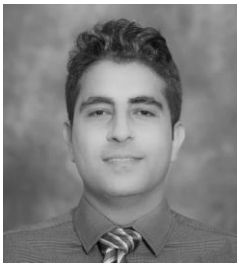

***Nozhan Hosseini*** is a PhD Candidate at the University of South Carolina Electrical Engineering department. He has over 5 years of experience in wireless and communication sciences. He received his M.Sc. from University of Bologna in 2015. His primary area of interest and expertise are signal processing, wireless channel characterization, physical layer design, Orthogonal chirp spread spectrum and software defined radios (SDR).

***Hosseinali Jamal*** received his B.Sc. in Electrical Engineering from Tafresh University, Iran, in 2009, and his M.Sc. in communication systems from Shahid Beheshti University, Iran, in 2013. And then he received his Ph.D. from University of South Carolina, USA, in Dec 2017. Since January 2018 he has been working as post-doc at University of South Carolina. His current research includes modulation and detection techniques for multicarrier communication systems (e.g. OFDM and FBMC). Modeling and simulating the communication links, channels and waveforms for air to ground communication systems.

***David W. Matolak*** has over 20 years experience in communication system research, development, design, and deployment, with private companies, government institutions, and academia, including AT&T Bell Labs, L3 Communication Systems, MITRE, and Lockheed Martin. He received the B.S. degree from The Pennsylvania State University, University Park, PA, the M.S. degree from The University of Massachusetts, Amherst, MA, and the Ph.D. degree from The University of Virginia, Charlottesville, VA, all in electrical engineering. He has over 200 publications, eight patents, and expertise in wireless channel characterization, spread spectrum, ad hoc networking, and their application in civil and military terrestrial, aeronautical, and satellite communication systems. He has been a visiting professor at several institutions in the USA and Europe, has worked on multiple NASA aviation projects, and serves on the ITU Radio Propagation Study Group. He was with Ohio University from 1999-2012, and since 2012 has been a professor at the University of South Carolina. His research interests are radio channel modeling and communication techniques for non-stationary fading channels, multicarrier transmission, aviation communications, and mobile ad hoc networks. Prof. Matolak is a member of Eta Kappa Nu, Sigma Xi, Tau Beta Pi, URSI, ASEE, AIAA, and a senior member of IEEE.

***Jamal Haque*** works for Honeywell International Inc., Aerospace division, as Staff Scientist R&D. He received his B.S., M.S. and Ph. D. degrees in Electrical Engineering from the University of South Florida Tampa, Florida. Dr. Haque's research interests are wireless systems, OFDM-based systems in high mobile platforms, synchronization, channel estimation, cognitive software defined radio, channel coding, high-speed connectivity and robust space processing systems and architectures. Prior to Honeywell, he worked at advance development groups at AT&T, Rockwell and Lucent (Bell Labs) technology on voice band modem, xDSL modem and Sirius Satellite Radio. He has over sixteen years of Telecommunication and Aerospace products design and development experience in the area of communication and signal processing. He holds several US Patents

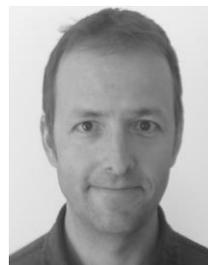

***Thomas Magesacher*** received a Ph.D. degree in Signal Processing in 2006 and a Docent degree in Telecommunications in 2011, both from Lund University, Sweden. During 2007/2008, he spent a year as postdoctoral fellow at Stanford University, CA. Since 2011, he holds on Associate Professor position at the Department of Electrical and Information Technology at Lund University. His research interests include applied *information processing and optimization for communications, channel modeling, transceiver architectures, and nonlinear signal processing for power-amplifier optimization. Since 2017, he is the principal system engineer at CesiumAstro Inc., a startup based in Austin, TX, developing integrated end-to-end communication systems for satellites, UAVs, and other space or airborne platforms.*